\documentstyle[prb,aps,psfig,multicol]{revtex} 

\begin{document}
\draft
\title{Current dependence of grain boundary magnetoresistance
in La$_{0.67}$Ca$_{0.33}$MnO$_3$ films}

\author{W.~Westerburg, F.~Martin, S.~Friedrich, M.~Maier, and G.~Jakob}

\address{Institut f\"ur Physik, Johannes Gutenberg-Universit\"at Mainz,
D-55099 Mainz, Germany}

\date{20 January 1999}

\maketitle

\begin{abstract}
We prepared epitaxial ferromagnetic manganite films on 45$^{\circ}$ bicrystal substrates by pulsed laser ablation.
Their low- and high-field magnetoresistance (MR) was measured as a function of magnetic field, temperature and
current. At low temperatures hysteretic changes in resistivity up to 70\,\% due to switching of magnetic domains
at the coercitive field are observed. The strongly non-ohmic behavior of the current-voltage ($I$-$V$) leads to a
complete suppression of the MR effect at high bias currents with the identical current dependence at low and high
magnetic fields. We discuss the data in view of tunneling and mesoscale magnetic transport models and propose an
explicit dependence of the spin polarization on the applied current in the grain boundary region.
\end{abstract}
%
%
%
%

%
\begin{multicols}{2}
\section{ Introduction}
Colossal magnetoresistance (CMR) in thin films of manganese based perovskites has been investigated intensively
during the last years. One motivation is the prospect to use these materials for magnetic devices, but a strong
field in the Tesla range is necessary to obtain large resistance changes.
\\
Another peculiarity, the highly spin polarized conduction band, allows to achieve high MR in low fields, because
of the spin polarized tunneling of electrons between grains \cite{Hwang96}. This can be realized by
heteroepitaxial tunnel junctions with electrodes of manganites and a thin insulating barrier \cite{Li97,Viret97}.
Other approaches are to produce polycrystalline films \cite{Gupta96}, bulk materials \cite{Hwang96}, ramp-edge
junctions \cite{Kwon98}, or thin films which are under compressive strain \cite{Wang98}. As is well known from the
high-$T_C$ superconductors, a further access is to deposit thin manganite films on bicrystal substrates to realize
a single artificial grain boundary (GB) \cite{Mathur97,Steenbeck98}.
\\
In order to understand the mechanism of the large low-field MR, we have carefully studied the resistivity of thin
manganite films with and without an artificial GB, induced by a bicrystal substrate, as a function of temperature,
magnetic field and current.

\section{ Experimental}
Epitaxial thin films of La$_{0.67}$Ca$_{0.33}$MnO$_3$ (LCMO) were grown from a stoichiometric target by pulsed
laser deposition (KrF Laser, $\lambda=248$\,nm). As substrates we used 45$^{\circ}$ SrTiO$_3$ bicrystals or
monocrystals. The optimized deposition conditions were a substrate temperature of 950$^{\circ}$C in an oxygen
partial pressure of 14\,Pa. The deposition rate was 0.3\,\AA{}/pulse with a pulse frequency of 3\,Hz. With a
Mireau interferometer the film thickness was determined to 190\,nm. After deposition the samples cooled in
500\,hPa O$_2$ atmosphere without further annealing.
\\
In X-ray diffraction in Bragg-Brentano geometry only film reflections corresponding to a (0~0~$l$) orientation of
the cubic perovskite cell are visible. The lattice parameter normal to the substrate plane  ($c=3.82$\,\AA) is
smaller than the bulk value and indicates a tensile strain of the $a,b$ axis due to the lattice mismatch of the
substrate. Rocking angle analysis shows epitaxial $c$-axis oriented growth with an angular spread smaller than
0.03$^{\circ}$ on both sides of the artificial GB. The in-plane orientation was studied by $\phi$-scans on both
sides of the bicrystal using the group of symmetry equivalent (2~0~2) reflections. The cubic perovskite axes of
the film are parallel to that of the SrTiO$_3$ bicrystal substrate with an angular spread in the orientation of
the individual epitaxial grains of less than 0.5$^{\circ}$. Thus the 45$^{\circ}$ GB in the film induced by the
bicrystal substrate is the only relevant large-angle GB in the following experiment.
\\

The surface morphology of the samples was studied by scanning electron (SEM) and by atomic force microscopy (AFM).
A shallow rectangular trench structure of 200\,nm width and 5\,nm depth along the grain boundary was visible in
AFM. The GB itself was identified as a further sharp dip with characteristic dimensions close to the resolution
limit of the used wide area scanner. The SEM scan in Fig.~1 shows that the actual distorted GB region is smaller
than 50\,nm.
\\
The films were patterned with conventional photolithography and chemically etched in an acidic solution of
hydrogen peroxide. We used a meander track which crosses the GB eleven times. The total length is 5.6\,mm and the
width is 0.1\,mm. For comparison an identical second structure was prepared adjacent to the GB, but not crossing
it.
\\
The resistivity was measured in a superconducting magnet cryostat by standard four-point technique with a constant
DC current between 10\,nA and 100\,$\mu$A. For higher currents up to 10\,mA current pulses of 3\,ms length were
applied in order to minimize sample heating. To see the low-field MR it is necessary to diminish demagnetization
effects, which requires that the film plane is parallel to
 the magnetic field $H$. Further the angle between the GB
and the magnetic field is important for the MR ratio. In this paper we show the results for $H$ parallel GB, where
the low-field MR is largest.

\section{ Results and Discussion}
The resistances of the meander tracks as a function of temperature show the usual maximum in the resistivity and
its suppression in high magnetic fields. The conductivity above the Curie-temperature $T_C=225$\,K is thermally
activated, which can be explained by small polaron hopping \cite{Jakob98_2}. At low temperatures the behavior of
the two meanders is not the same. The difference between the meander resistance on and beside the GB isolates the
temperature dependence of the true GB resistance. In agreement with published data there is a broad flat peak
around 100\,K below $T_C$ \cite{Isaac98}.
\\
In Fig.~2 the resistivity as a function of magnetic field at 4.2\,K measured with a constant DC current of
1\,$\mu$A is shown. Here the magnetic field is parallel to the plane of the film and to the GB. The MR defined as
$\Delta{}R/R=[R(H)-R(H=0)]/R(H=0)$ reaches its maximum value of 70\,\% in a field of 120\,Oe at 4.2\,K, which is
approximately equal to the coercitive field. Assuming spin polarized tunneling \cite{Julliere75}, the maximum MR
is given by
\begin{equation}
\Delta{}R/R_{max}=(R_{\uparrow{}\downarrow}-R_{\uparrow{}\uparrow})/R_{\uparrow{}\uparrow}=2P^2/(1-P^2)
\end{equation}
with $P=(n_{\uparrow}-n_{\downarrow})/(n_{\uparrow}+n_{\downarrow})$ the spin polarization parameter and
$R_{\uparrow{}\uparrow}$ and $R_{\uparrow{}\downarrow}$ the resistances for the parallel and antiparallel
configurations of the magnetizations, respectively. A value of 70\,\% for $\Delta{}R/R$ will result in
$P\approx0.51$ for LCMO. We have to mention here that the different steps and switches in Fig.~2 indicate that in
the neighborhood of the GB multiple magnetic domains can exist and thus the spin polarization parameter will be
underestimated.
\\
With higher temperatures the maximum low-field MR decreases almost linearly down to 125\,K. At $T_C$ of the sample
the hysteretic low-field MR vanishes, as can be seen in Fig.~3. This temperature dependence is different to that
of magnetic tunnel junctions \cite{Li97}. The magnetic field value, where the maximum MR ratio occurs, does not
shift with temperature up to 200\,K. The inset of Fig.~3 shows a hysteresis loop measured close to $T_C$
($T=200$\,K). At higher temperature the hysteretic low-field MR becomes less pronounced and the intrinsic negative
CMR effect becomes visible. Above 150\,K the meander track not passing the GB shows comparable MR effects. Thus
the artificial GB acts in this case as a magnetic field independent series resistor. This indicates loss of
magnetic order in the distorted GB region well below the bulk $T_C$ \cite{Evetts98}. The low-field MR visible
above 150\,K in both meanders shows a maximum value of 3\,\% at 175\,K. We think that this MR results from a large
number of small angle grain boundaries between the individual epitaxial grains. This
effect decreases at higher temperatures due to the loss of spin orientation in the individual grains near $T_C$ of
the sample. Below 175\,K the increasing magnetic domain size will lead to ferromagnetic coupling of a large number
of grains in each magnetic domain. Thus only grains at magnetic domain boundaries contribute to the low-field MR.
If at low temperatures the magnetic domains are much larger than the crystal grain size, the number of not spin
aligned grains is negligible and the low-field MR vanishes, as can be seen from the lower curve of Fig.~3. At
4.2\,K the noise level of the experiment sets an upper limit for the MR of the reference meander of $<0.05$\,\%.
Within the above description this requires a growth of the magnetic domain size by two orders of magnitude.
\\
The influence of the GB on the transport properties is also visible in the current-voltage ($I$-$V$) curves of the
meanders. In the paramagnetic regime both meanders show a nearly identical ohmic response. Below $T_C$ the one
which crosses the GB is strongly nonlinear in contrast to the ohmic behavior of the reference meander as can be
seen in Fig.~4 for $T=4.2$\,K in zero field. With increasing bias current the differential conductivity of the GB
meander becomes asymptotic to that of the reference meander. Significant heating effects can be excluded, because
the resistance decreases with increasing current, while the temperature coefficient of the resistivity is
positive.
\\
Nonlinear $I$-$V$ curves are often taken as an indication for tunneling transport. In ferromagnetic materials the
inclusion of spin dependent density of states factors and hysteretic magnetizations on both sides of the tunneling
barrier can explain a hysteretic MR. A description of the nonlinear $I$-$V$ curve shown in Fig.~4 with the
standard Simmons tunneling model in the parabolic approximation \cite{Simmons63} is able to reproduce the data
within the linewidth of the figure. In addition to the tunneling parameters one has to take into account in a
fitting procedure the resistivity drop in the non distorted regions. This can be done either as a free parameter
or by using the resistivity of the reference meander. We obtained as minimizing parameters a barrier height of
$\Phi=0.26$\,eV, a width of 2.1\,nm, and a resistance of the undistorted region of 1630\,$\Omega$  (reference
meander $R=1770$\,$\Omega$). Such a small width of the barrier is not expected in view of the clear visibility of
the GB in the SEM scan. In spite of the nearly perfect fit possible for the nonlinear $I$-$V$ curve of Fig.~4, the
investigation of direct $I$-$V$ data gives vanishing weight to the true differential conductivity at zero bias
current. Therefore it is necessary to inspect the differential conductivity shown in Fig.~5. For very low bias
voltages a nearly parabolic behavior of the differential conductivity, as predicted by the Simmons model, was
found by Steenbeck {\it et al.} \cite{Steenbeck98}. However, the inclusion of high bias current data in our
measurements results in stronger systematic deviations of a Simmons model from the experimental data at low
voltages. The differential conductivity sketched in Fig. 5 was calculated in the parabolic approximation. In view
of the low barrier energy this is not appropriate at very high bias currents, which correspond to high bias
voltages. But using the more general formulas given by Simmons, will even enhance the deviations.
Thus it is unclear if charge transport takes place by GB-tunneling.
\\
Also the observed anisotropy of the MR with respect to the magnetic field direction cannot be easily
understood in this framework. Therefore as an alternative mechanism Evetts {\it et al.} \cite{Evetts98} proposed,
that the GB does not act as a tunneling barrier, but as a mesoscale region with distorted magnetic and transport
properties. The resistivity of the GB region is assumed to be a function of the GB magnetization, which depends on
the effective field acting in the GB region. This interpretation is in agreement with the observed anisotropy.
Also the loss of magnetic order in the mesoscale region can naturally occur well below the bulk $T_C$ and will be
responsible for the vanishing low-field MR at higher temperatures. The up to now proposed transport mechanisms in
the mesoscale region, however, are essentially thermally
activated and ohmic. Therefore they cannot simply account for nonlinear $I$-$V$ curves.
\\
\\
A dependence of the hysteretic low-field MR on the current strength, which we present in the following, is in
neither in a tunneling model nor in a mesoscale transport model a priori expected.
\\
We investigated the dependence of the low-field MR on the transport current on a meander which shows a maximum MR
peak $\Delta{}R/R_{max}$ of 22\,\% at $T=4.2$\,K. The applied current range spanned six decades, from 10\,nA to
10\,mA. Fig.~6 shows that the width of the hysteresis loops is current independent, while there is a strong
current dependence of the height of the MR peaks. The suppression of the MR with current is shown in Fig.~7 for
bias currents from 22\,\% for 10\,nA to 1.4\,\% for 10\,mA.
\\

A close inspection of the data shows, that a current dependence exists also for magnetic field strengths well
above the hysteretic region. The decrease of this negative CMR of the GB meander with increasing current is shown
in an enlarged scale in Fig.~8. On the reference meander no current dependence of the resistivity is observed. The
negative linear CMR slope ${\rm d}R/{\rm d}(\mu_0 H)$ as function of the current was measured up to 10\,T and is
also plotted in Fig.~7. The close correlation between the CMR suppression and the low-field MR peak suppression
with current indicates the same physical origin for both effects in the GB region. Since high bias currents
correspond to high bias voltages one possible explanation is, that at high bias voltage a new spin independent
channel for charge transport opens. However, due to the intimate relationship between ferromagnetism and spin
polarized charge transport in the manganites a different scenario is possible: A polarized charge carrier from the
undisturbed region will induce due to double exchange a spin polarization in the distorted region during its
transition, i.e.\ the carrier leaves a trail of localized spins behind, which are more than usually aligned. Each
manganese ion will encounter the passage of up to $10^8$ polarized carriers per second for the current values used
in our experiment. If the spin relaxation time is much longer than 10$^{-8}$\,s the charge transport induced
magnetization will be nonzero in the time average and a second carrier can use this spin polarized path before the
spin relaxation has destroyed the induced order! The existence of electron spin resonance measurements at
microwave frequencies in the manganites \cite{shengelaya96}, shows that the relaxation time is longer than
(10\,GHz)$^{-1}$, and therefore this scenario is imaginable. In the limit of full magnetization of the distorted
region the GB meander resistivity will be indistinguishable from that of the reference meander. This
interpretation does not depend on the details of the transport of the charge carriers from one manganese ion to
the next in the mesoscale region, but requires only a hopping probability depending on the respective spin
alignment as proposed by several authors \cite{Jakob98_2,Coey95,Dionne96,Wagner98}.

\section{ Summary}
By pulsed laser ablation we prepared thin films of La$_{0.67}$Ca$_{0.33}$MnO$_3$ on SrTiO$_3$ bicrystal
substrates. X-ray diffraction showed full epitaxial growth on both sides of the bicrystal substrate. Their low-
and high-field MR was measured as a function of magnetic field, temperature and current. We obtained hysteretic MR
values up to 70\,\% at low temperatures in low magnetic fields. The MR is induced by switching of magnetic domains
at the coercitive field. With increasing temperature the artificial grain boundary MR decreases almost linearly.
We observed a highly non ohmic behavior of the current-voltage relation, which leads to a nearly complete
suppression of the MR effect at high bias currents. The current dependence of low- and high-field MR is identical,
which indicates the same physical origin for this suppression. We discussed the data in view of proposed tunneling
and mesoscale magnetic transport models. An explicit dependence of the spin polarization on the applied current is
necessary for a consistent interpretation. In the mesoscale transport model the number of
manganese ions in the distorted grain boundary region is comparable to the number of charge carriers passing the
region within the spin relaxation time and we propose a current induced magnetization change.

\acknowledgments
This work was supported by the Deutsche Forschungsgemeinschaft through project JA821/1-3.

\begin{figure}[t]
\caption{SEM scan of the GB induced in
the LCMO film. Film growth is distorted on the GB in a region with a width smaller than 50\,nm.
For a detailed picture: http://www.uni-mainz.de/$\sim$westerbu/MnBRD.htm} \label{Fig1}
\end{figure}

\begin{figure}[b]
\centerline{\psfig{file=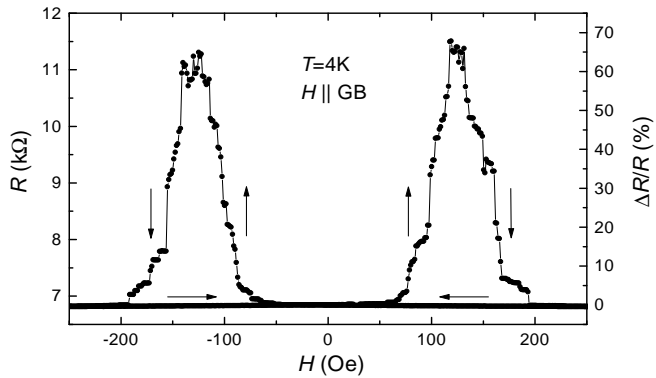,width=1.0\columnwidth}} \vspace{4.5em} \caption{GB resistance and MR at
$T=4.2$\,K as a function of magnetic field.} \label{Fig2}
\end{figure}

\begin{figure}[ht]
\centerline{\psfig{file=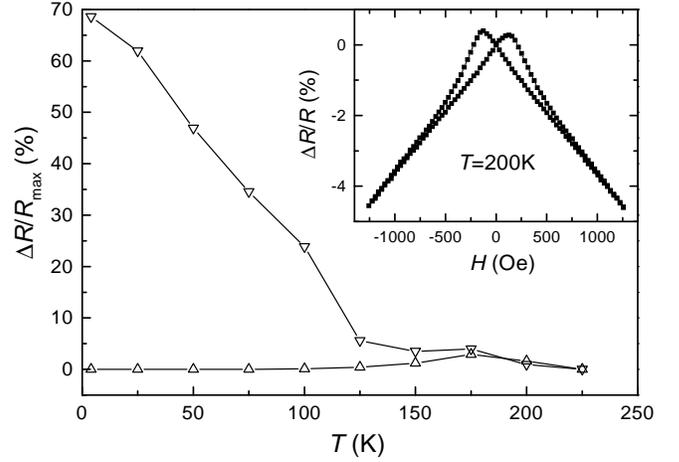,width=1.0\columnwidth}} \vspace{4.5em} \caption{Temperature dependence of the
maximum low-field MR for the meander crossing (down triangle) and not crossing (up triangle) the GB. The inset
shows the hysteretic MR behavior near $T_C$.} \label{Fig3}
\end{figure}

\begin{figure}[t]
\centerline{\psfig{file=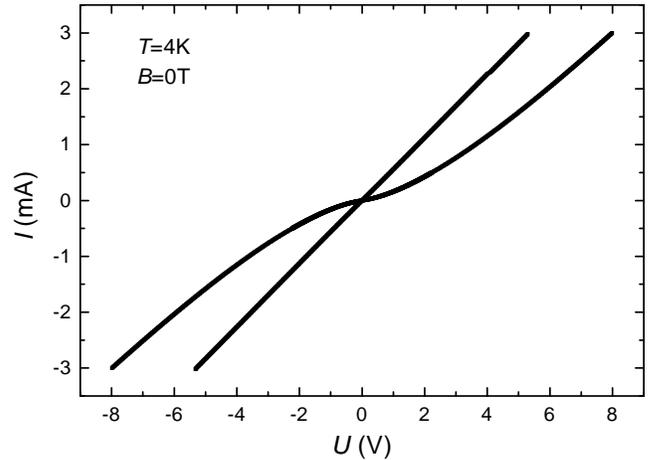,width=1.0\columnwidth}} \vspace{4.5em} \caption{$I$-$V$ curves of the GB 
(non-linear) and
reference meanders (linear) in zero field at $T=4.2$\,K.} \label{Fig4}
\end{figure}

\begin{figure}[h]
\centerline{\psfig{file=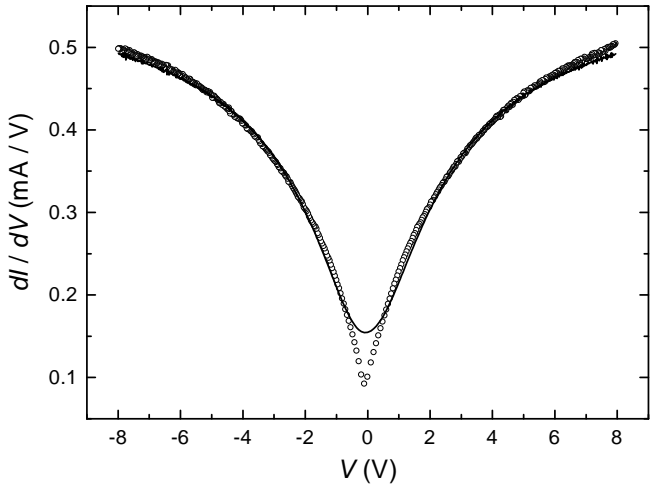,width=1.0\columnwidth}} \vspace{4.5em} \caption{Measured (symbols) and
calculated (line) differential conductivity of GB meander. The calculation includes a series resistor and a
tunneling element described by the Simmons model in the parabolic approximation.} \label{Fig5}
\end{figure}

\begin{figure}[h] \centerline{\psfig{file=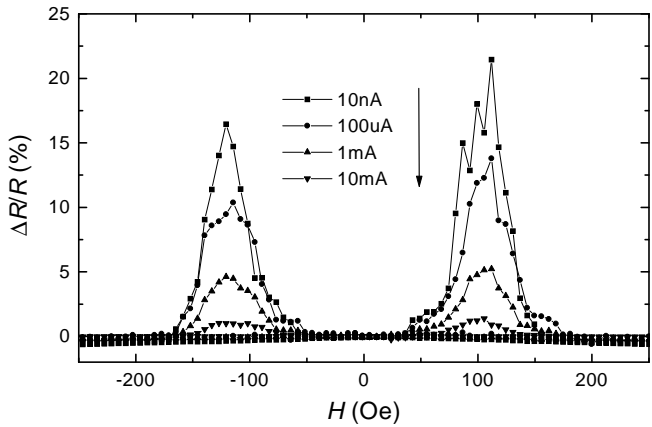,width=1.0\columnwidth}} \vspace{4.5em}
\caption{Hysteresis loops of the low-field MR measured with bias currents from 10\,nA to 10\,mA. For clarity not
all measured curves are shown.} \label{Fig6}
\end{figure}

\begin{figure}[h]
\centerline{\psfig{file=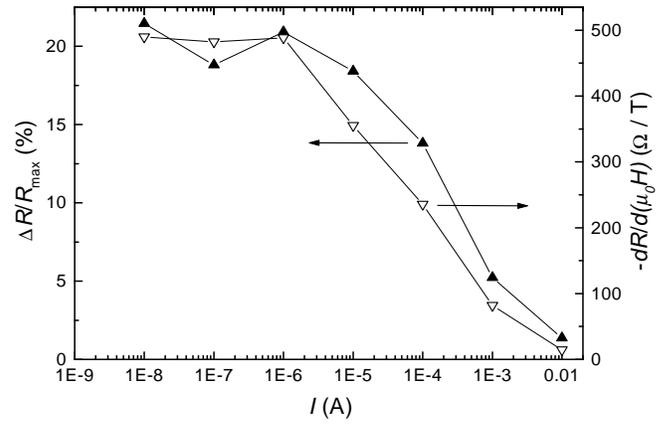,width=1.0\columnwidth}} \vspace{4.5em} \caption{Decrease of the low-field MR
at the GB with increasing applied current (left axis) and of the respective negative linear high-field CMR slope
${\rm d}R/{\rm d}(\mu_0 H)$  (right axis).} \label{Fig7}
\end{figure}

\begin{figure}[h]
\centerline{\psfig{file=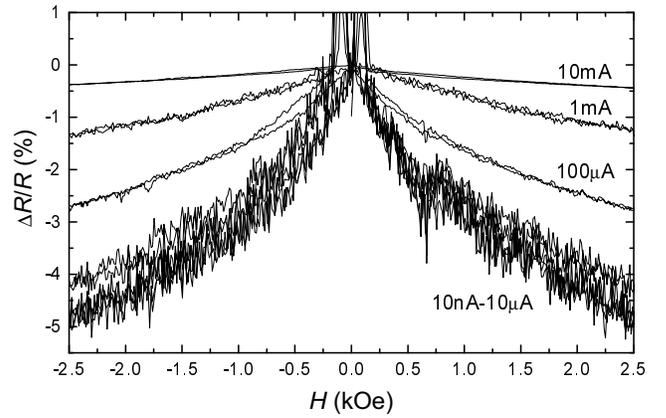,width=1.0\columnwidth}} \vspace{4.5em} \caption{Reversible colossal
magnetoresistance of the GB meander as function of applied field. The curves are measured with bias currents from
10\,nA to 10\,mA. The stronger positive low-field MR shown in Fig.~6 leads to the apparent divergence of the data
near $H=0$ in the scaling used here.} \label{Fig8}
\end{figure}
\end{multicols}
\end{document}